
\input harvmac

%

\def\bar{\overline}
\def\hat{\widehat}
\def\*{\star}
\def\({\left(}		
\def\){\right)}		
\def\[{\left[}		\def\BBL{\Bigl[}
\def\]{\right]}		\def\BBR{\Bigr]}
\def\lb{\[}
\def\rb{\]}
%
%
\def\frac#1#2{{#1 \over #2}}		
\def\inv#1{{1 \over #1}}

\def\d{\partial}

\def\ket#1{ | #1 \rangle}
\def\bra#1{ \langle #1 |}

\def\2pi{\hbox{$2\pi i$}}

\def\dsl{\raise.15ex\hbox{/}\kern-.57em\partial}
\def\Dsl{\,\raise.15ex\hbox{/}\mkern-.13.5mu D}
%
%
\def\th{\theta}		
\def\ga{\gamma}

\def\ep{\epsilon}
\def\la{\lambda}	
\def\de{\delta}		\def\De{\Delta}
		\def\Om{\Omega}
\def\sig{\sigma}	\def\Sig{\Sigma}
\def\vphi{\varphi}
%
%

%
%
\font\numbers=cmss12
\font\upright=cmu10 scaled\magstep1
\def\stroke{\vrule height8pt width0.4pt depth-0.1pt}
\def\topfleck{\vrule height8pt width0.5pt depth-5.9pt}
\def\botfleck{\vrule height2pt width0.5pt depth0.1pt}
\def\Zmath{\vcenter{\hbox{\numbers\rlap{\rlap{Z}\kern 0.8pt\topfleck}
	          \kern 2.2pt \rlap Z\kern 6pt\botfleck\kern 1pt}}}
\def\Qmath{\vcenter{\hbox{\upright\rlap{\rlap{Q}\kern
                   3.8pt\stroke}\phantom{Q}}}}
\def\Nmath{\vcenter{\hbox{\upright\rlap{I}\kern 1.7pt N}}}
\def\Cmath{\vcenter{\hbox{\upright\rlap{\rlap{C}\kern
                   3.8pt\stroke}\phantom{C}}}}
\def\Rmath{\vcenter{\hbox{\upright\rlap{I}\kern 1.7pt R}}}
\def\Z{\ifmmode\Zmath\else$\Zmath$\fi}
\def\Q{\ifmmode\Qmath\else$\Qmath$\fi}
\def\N{\ifmmode\Nmath\else$\Nmath$\fi}
\def\C{\ifmmode\Cmath\else$\Cmath$\fi}
\def\R{\ifmmode\Rmath\else$\Rmath$\fi}
\def\te{ \theta }

\Title{SPhT- 93-006}{\vbox{
\centerline{Yang-Baxter Equation in}
\centerline{~}
\centerline{Spin Chains with Long Range Interactions}}}
\vskip 1.5 cm
\centerline{ D. Bernard ${}^{(a)}$, M. Gaudin ${}^{(a)}$,
F.D.M. Haldane ${}^{(b)}$ and V. Pasquier ${}^{(a)}$ }
 \bigskip
${}^{(a)}$ Service de Physique Th\'eorique de Saclay,
\foot{Laboratoire de la Direction de Sciences de la Mati\`ere
du Commissariat \`a l'\'energie atomique.}
91191 Gif-sur-Yvette, France.

${}^{(b)}$ Department of Physics, Princeton University, Princeton NJ 08544,
USA.
\vskip 1.5 cm
Abstract

We consider the $su(n)$ spin chains with long range interactions
and the spin generalization of the Calogero-Sutherland models.
We show that their properties derive from a
transfer matrix obeying the Yang-Baxter equation.
We obtain the expression of the conserved quantities and
we diagonalise them.

\Date{01/93}

\newsec{Introduction.}
The most remarkable properties of the XXX chain with long range interactions
\ref\ha{F.D.M. Haldane,
Phys.Rev.Lett. 60 (1988) 635.}
\ref\sha{B.S. Shastry, Phys.Rev.Lett. 60 (1988) 639 }
are that its spectrum is additive and that the states are created
by filling a ``Dirac sea'' with particles obeying a
 ``Generalised Pauli principle"
\ref\habis{F.D.M. Haldane, Phys.Rev.Lett. 67 (1991) 937}.
Recently, it has become apparent
\ref\nous{F.D.M.Haldane, Z.N.C.Ha, J.C.Talstra,
D.Bernard and V.Pasquier, Phys.Rev.Lett.
69 (1992) 2021.}
that  the algebras underlying the symmetries of these models are
the Yangians \ref\drin{V.G.Drinfel'd, ``{\it Quantum Groups}",
Proc. of the ICM, Berkeley (1987) 798}.
In \nous, the first generators of the Yangian had been obtained,
 and the aim of this paper is to display the
full algebra.
To characterise it, we have constructed a transfer matrix
which satisfies the exchange relations\ref\skly{E.K.Sklyanin,
Funct.Anal.Appl. 16 (1983) 263} resulting
from the Yang-Baxter\ref\ba{R.J.Baxter, ``{\it Exactly solved models
in statistical mechanics}" Academic, London (1982)} equation,
(often called the $RLL=LLR$ relations\ref\fad{L.D. Faddeev,
``{\it Integrable models in 1+1 dimensional quantum field theory}"
Les Houches lectures, Elsevier Science Publishers (1984)}).
In the limit of infinite separation of the sites,
the transfer matrix reduces to the usual XXX chain transfer
matrix\ba.
In proving the exchange relations, the differentials operators
defined in \ref\pol{A.P.Polychronakos, Phys.Rev.Lett.  69 (1992) 703}
\ref\brin{L.Brink, T.H.Hansson and M.Vasiliev, Phys.Letters B286
(1992) 109. }
have turned out to be essential tools.

In the course of this work, we have been led to
consider\foot{This came through discussions with M. Douglas.}
models for which the lattice sites
are replaced by dynamical particles, see also \ref\vrac{
J.Minahan and A. Polychronakos, CU-TP-566 (1992) preprint \semi
K.Hikami and M.Wadati, preprint (1992). }.
They are the spin generalisations of the
Calogero-Sutherland models\ref\cal{P.Calogero, J. Math. Phys.
10 (1969) 2191  \semi P. Calogero, O. Ragnisco and C. Marchioro,
Lett. Nuevo Cim.  13 (1975) 383. }
\ref\su{B.Sutherland, Phys.Rev. A5 (1972) 1372, and
reference therin. }.
In these models, there also exists a transfer matrix
obeying the Yang-Baxter equation.
However, an important difference between the
two situations is that the
transfer matrix of the dynamical models
 always commutes with the Hamiltonian,
whereas in the lattice model case, it commutes
only if the lattice is translation invariant.
The generating function for the Hamiltonians
is not given by the trace of the transfer
matrix, because this trace does
not commute with the Yangian.
In  the dynamical case
it is given by the quantum determinant.
In the lattice case, the determinant is a
c-number which contains enough information to
recover the spectrum.

In part two we define the dynamical models
and we present their Lax pair.
In part three we prove the Yang-Baxter equation
for the transfer matrix.
In part four, we derive the conserved quantities
and we obtain their eigenvalues.
Finally, we consider the spin models and we
decompose the $su(2)$-spin chain into
 irreducible Yangian representations.

\newsec{The dynamical models.}

The dynamical models are $su(p)$ generalisations of the
Calogero-Sutherland model. There are $M$ particles interacting by long range
forces. Their positions are parameterised by complex numbers $z_i$,
$i=1,\cdots,M$, and each particle carries a spin $\sigma$
 with $p$ possible values.
If the particles are on the unit circle, we take $z_j=\exp(i\nu_j)$,
and if they are on the line we take $z_j=\exp\nu_j$, with $\nu_j$ real.
Their dynamics are governed by the Hamiltonian :
\eqn\Iiii{ H_D = \sum_{j=1}^M (z_j\d_{z_j})^2 + \sum_{i\not= j}
	\la(P_{ij}+\la) h_{ij} }
where $\la$ is a coupling constant and $P_{ij}$ exchanges the spins
of the particles $i$ and $j$. The function $h_{ij}$ is :
\eqn\Iii{ h_{ij} = \frac { z_i z_j}{ (z_i-z_j)(z_j-z_i)} }

Integrability is guaranteed by the existence of a Lax pair.
It consists of
 two matrices $L_{ij}$ and $M_{ij}$
with operator entries obeying:
\eqn\IIi{ \BBL\ H_D\ ,\ L_{ij}\ \BBR = \sum_k \Bigl({
L_{ik} M_{kj}\ - \ M_{ik}L_{kj} }\Bigr) }
A possible choice is given by:
\eqn\IIii{\eqalign{
L_{ij} &= \de_{ij} z_j\d_{z_j} +
(1-\de_{ij})\la \te_{ij}P_{ij}\cr
M_{ij} &= -\de_{ij}2\la \sum_k( h_{ik} P_{ik}) +2\la(1-\de_{ij})
 h_{ij} P_{ij} \cr}}
with:
\eqn\IIiv{
\te_{ij}=\frac{z_i}{z_i-z_j} }

We denote by $X_j^{ab}$, $a,b=1,\cdots,p$, the
matrices which act as $\ket{a}\bra{b}$  on the spins
of the $j^{th}$ particle and leave the other particles untouched.
Using equation \IIi~ and the fact that
$ \lb H_D , X_{j}^{ab} \rb = \sum_k
X_{k}^{ab} M_{kj}$
and  $ \sum_{j} M_{ij}=0 $ , one deduces that the
quantities defined by:
\eqn\IIIv{ T_n^{ab}\ =\
\sum_{ij}X_{i}^{ab}(L^{n})_{ij} }
commute with the Hamiltonian $H_D$.
Here $L^n$ denotes the $n^{th}$ power of the matrix $L_{ij}$.
 Since the $T_n^{ab}$ do not commute with each other, the
 spectrum is degenerate.
We study their algebra
in the next section.

\newsec{The transfer matrices.}
In order to arrange the algebra of the $T_n^{ab}$'s,
we introduce the transfer matrix $T(u)$ obeying the Yang-Baxter equation:
\eqn\IVi{
R_{00'}(u-v)\ T^0(u)\ T^{0'}(v)\ =\
T^{0'}(v)\ T^0(u)\ R_{00'}(u-v)}
According to standard notations \fad , $T^0(u)$ denotes the matrix
$T(u)\otimes 1$ and $T^{0'}(v)$ the matrix $1\otimes T(v)$.
The matrix $R(u)$ is the solution of the Yang-Baxter equation given by:
\eqn\IVii{
R(u)\ =\ u + \la P_{00'} }
where $P_{00'}$ is the permutation operator which exchanges the two auxiliary
spaces $0$ and $0'$. Equation \IVi~ expresses the non-commutativity of
the operator matrix elements of $T(u)$. The expression which we obtain
for $T^0(u)$ is given by:
\eqn\IViii{
T^0(u)\ =\ 1 + \la \sum_{i,j=1}^M P_{0i}\({\inv{u-L}}\)_{ij} }
where $L$ is the matrix
defined in eq. \IIii.
If we set $T^0(u)= \sum_{a,b=1}^p X^0_{ba}T^{ab}(u)$, and expand it in
powers of $\inv{u}$, we find:
\eqn\IVv{
T^{ab}(u)\ =\ \de^{ab} +
 \sum_{n=0}^\infty\frac{\la}{u^{n+1}} T_n^{ab}}
where the $T^{ab}_n$'s have been defined in eq. \IIIv .
We can motivate this expression as follows: It commutes with the Hamiltonian
$H_D$, the $\inv{u^2}$ coefficient coincides with (a slight generalization of)
the generators of the Yangians identified in \nous, and, as we will
discuss in a next section, in a specific limit it gives back the
transfer matrix of an XXX chain.
When the $z_i$ which define the functions $\theta_{ij}$
in equation \IIiv~ are complex numbers of modulus one,
an important property of this transfer matrix
is its hermiticity: $T^{ab\dagger}(u)=T^{ba}(\bar u)$.
As a result, the induced Yangian representation is fully
decomposable into a direct sum of irreducible representations.
\bigskip
In order to prove \IVi, we use the formalism introduced in \pol\brin .
Let us define the three permutation groups: $\Sigma_1,\ \Sigma_2$
and $\Sigma_3$ respectively generated by $K_{ij},\ P_{ij}$ and the product
$(P_{ij}K_{ij})$. The operator $K_{ij}$ exchanges the positions $z_i$,
and $P_{ij}$ exchanges the spins at positions $i$ and $j$: i.e.
$K_{ij}z_j=z_iK_{ij}$ and $P_{ij}\sigma_j=\sigma_iP_{ij}$.
We define the differentials:
\eqn\IVwi{
D_i= z_i\d_{z_i} + \la \sum_{i\not= j} \theta_{ij} K_{ij} }
They obey the relations:
\eqn\IVw{\eqalign{
K_{ij}\ D_i\  &=\ D_j\  K_{ij} \cr
\BBL\ D_i\ ,\ D_j\ \BBR\ &= \la(D_i-D_j)\ K_{ij} \cr}}
We also define a projection $\pi$  from the cross product $\Sigma_2
\triangleright \Sigma_3$ into $\Sigma_2$ by $\pi(\sigma \sigma')=\sigma$
for $\sigma\in\Sigma_2$ and $\sigma'\in\Sigma_3$.
We use it to eliminate the permutations of $\Sigma_1$ by replacing them
with those of $\Sigma_2$.
For example: $\pi(K_{12})=\pi(P_{12}(P_{12}K_{12}))=P_{12}$.
One can view this projection as the result of acting on
symmetric functions under permutations in $\Sigma_3$. The expression
\IViii~ of the transfer matrix then takes the following form:
\eqn\IVvi{
T^0(u)\ =\ \pi\Bigl({ 1 + \la \sum_{i=1}^M \frac{P_{0i}}{u-D_i}}\Bigr) }
which results from $\pi(D^n_i)=\sum_j L^n_{ij}$. Since $T^0(u)$ is
invariant under simultaneous permutations of the spins and the coordinates,
a product of projections can be replaced by the projection of the
product. Therefore, we omit the symbol $\pi$ and set equal to one
any permutation of $\Sigma_3$ appearing to the right of an expression.
Equation \IVi~ then recasts into the form :
\eqn\IVviii{\eqalign{
&(u-v+\la P_{00'}) \Bigl({ \sum_{i=1}^M
(1+\frac{\la P_{0i}}{u-D_i}) (1+\frac{\la P_{0'i}}{v-D_i}})
+\sum_{i\not=j} (\frac{\la P_{0i}}{u-D_i})(\frac{\la P_{0'j}}{v-D_j} )\Bigr)\cr
=& \Bigl({ \sum_{i=1}^M(1+\frac{\la P_{0'i}}{v-D_i}) (1+\frac{\la
P_{0i}}{u-D_i})
+\sum_{i\not=j}(\frac{\la P_{0'j}}{v-D_j})(\frac{\la P_{0i}}{u-D_i})}\Bigl)
 (u-v+\la P_{00'}) \cr}}
First consider the sum over $i$:
 each term in the sum is a Yang-Baxter equation
for an elementary transfer matrix, $T_i(u)=1+\frac{\la P_{oi}}{u-D_i}$,
 with a spectral parameter equal to $u-D_i$.
The equality is therefore satisfied by each term independently.
 Then, consider the sum with $i\not=j$. Using the identities:
\eqn\IVix{\eqalign{
P_{00'} P_{0i} P_{0'j}\ &=\ K_{ij} P_{0i} P_{0'j} (K_{ij}P_{ij})\cr
P_{0i} P_{0'j} P_{00'}\ &=\ K_{ij} P_{0'i} P_{0j} (K_{ij}P_{ij})\cr}}
the remaining terms reduce to :
\eqn\IVx{
 (u-v+\la K_{ij}) \({\inv{v-D_i}}\) \({\inv{u-D_j}}\)
=\ \({ \inv{u-D_j}}\)\({ \inv{v-D_i}}\) (u-v+\la K_{ij})}
or equivalently,
\eqn\IVxi{
 (u-v+\la K_{ij}) (u-D_j) (v-D_i)\ =\ (v-D_i) (u-D_j) (u-v+\la K_{ij}) }
This results from eq. \IVw. It proves the Yang-Baxter equation \IVi~ for
the transfer matrix given in eq. \IViii.

\bigskip
Finally, let us discuss the rational limit
of the preceeding results.
If we set $z_j=\exp(\ga\nu_j)$
and let $\ga$ go to zero, the Hamiltonian becomes:
\eqn\Zi{
H_D = -\sum_{j=1}^M (\d_{\nu_j})^2 + \sum_{i\not= j}
	\frac{\la(P_{ij}+\la)}{(\nu_i-\nu_j)^2} }
In order to reach this limit, we need to rescale
$D_j$ and $\la$ by $\ga$. When $\ga$ goes to zero,
the $T_n^{ab}$ algebra degenerates to:
\eqn\Zii{
\BBL T_n^{ab}, T_m^{cd} \BBR
= \de_{bc} T_{n+m}^{ad} - \de_{ad} T_{n+m}^{cb} }
In that case, the traces $\sum_a T_n^{aa}$ define the conserved
quantities which commute with the $ T_n^{ab} $.

\newsec{The conserved quantities.}
One usually generates the conserved quantities
by means of the trace of the transfer matrix.
 This defines quantities which commute among themselves
 but not with the $T_n^{ab}$.
Here, in order to satisfy this condition, we are led to take
the quantum determinant as the generating function.
It is given by  \skly \drin :
\eqn\IVxii{
Det_q\ T(u)\ =\ \sum_{\sig \in\Sig_p} \ep(\sig)
T_{1 \sig_1 }(u-(p-1)\la) T_{2 \sig_2} (u-(p-2)\la)\cdots
T_{p \sig_p }(u)}
The first two
 non-trivial terms produce the momentum and the Hamiltonian:
\eqn\IVxiii{\eqalign{
Det_q T(u)\ =\ 1 &+ \frac{\la}{u} M +
\frac{\la}{u^2}\Bigl({\sum_{j=1}^M z_j\d_{z_j} + \la \frac{M(M-1)}{2}}\Bigr)
\cr
 &+ \frac{\la}{u^3}\Bigl({ H_D+ \la(M-1)\sum_{j=1}^M z_j\d_{z_j}
+ \la^2 \frac{M(M-1)(M-2)}{6}}\Bigr)
+\cdots \cr} }
\bigskip
We now give an alternative description of the transfer matrix
which makes the actual computation of the quantum determinant
much simpler.
 Let us modify the differentials $D_i$ introduced in \IVwi~
 as follows:
\eqn\Wi{
\eqalign{
\hat D_i &=D_i - \la \sum_{j<i} K_{ij} \cr
&= z_i\d_{z_i} + \la \sum_{j>i} \theta_{ij} K_{ij}
- \la \sum_{j<i} \theta_{ji} K_{ij} \cr} }
Unlike the $D_i$, the differentials $\hat D_i$ commute:
\eqn\Wii{
\BBL\ \hat D_i\ ,\ \hat D_j\ \BBR\ = 0 }
In term of these differentials, we can define a transfer
matrix which obviously satisfies the Yang-Baxter equation.
It is given by:
\eqn\Wiii{
\hat T^0(u) =
\Bigl({ 1 +\frac{\la P_{01}}{u-\hat D_1} }\Bigr)
\Bigl({ 1 +\frac{\la P_{02}}{u-\hat D_2} }\Bigr) \cdots
\Bigl({ 1 +\frac{\la P_{0M}}{u-\hat D_M} }\Bigr) }
The right hand side of \Wiii~ is the co-product of
elementary transfer matrices $\hat T_i$
with their spectral parameters
equal to $u-\hat D_i$.
It is not difficult to show that
 the projection of
 $\hat T^0$ by $\pi$,
$\pi(\hat T ^0)$,
is invariant under $\Sigma_3$.
This imposes the  choice of the coefficients
 of $P_{0i}$ to be $\la$ in equation \Wiii.
 Thus, when we apply the
projection $\pi$ to the Yang-Baxter equation \IVi, we can
replace the projection of the product
$\hat T^0 \hat T^{0'}$ by the product of
the projections. Therefore,
$\pi(\hat T ^0)$ also satisfies the Yang-Baxter equation.
Indeed, one can check that it coincides with the definition
of $T^0$ in \IVvi.

 The quantum determinant of $\hat T$
is the product of the quantum determinants of the $\hat T_i$.
 It is equal to:
\eqn\Wiv{
Det_q\hat T(u)=\ \frac{\hat \Delta_M(u+\la)}{\hat \Delta_M(u)}}
where:
\eqn\Wv{
\hat \Delta_M(u)=\prod_{i=1}^M(u-\hat D_i)}
Using a similar argument to that outlined above,
the invariance of $\hat \Delta_M(u)$
under $\Sigma_1$
allows its replacement by its projection
$\Delta_M(u)$ in \Wiv~ to
get the quantum
determinant of $T(u)$.
Note that the polynomial transfer matrix $\Delta _M(u) T(u)$
also satisfies the Yang-Baxter equation.

The coefficients $C_p$ of the polynomial $\Delta_M(u)$
define a complete set of commuting operators.
We conjecture that they
are given by the following ``Wick'' expansion:
\eqn\Wvi{
\eqalign{
C_p &=
 \sum_{i_1<\cdots <i_p}
\pi(\hat D_{i_1}\hat D_{i_2} \cdots\hat D_{i_p}) \cr
&=
 \sum_{i_1<\cdots <i_p}\
\sum_{I\cup J=\lbrace i_1,\cdots ,i_p\rbrace}
F_J\
\prod_{i\in I}z_i\d_{z_i} \cr}}
with:
\eqn\Wvii{
F_J= \sum_{\cup \lbrace j_k,j'_k\rbrace =J}
\prod_k\({-\la (\la+P_{j_kj'_k})h_{j_k j'_k} }\)  }
In \Wvi~ and \Wvii, the symbol $\cup$ stands for the
union of sets which do not intersect.
  $F_J$ is equal one if $J=\emptyset$ and
to zero if the number of elements
of J is odd. The functions
$h_{ij}$ are defined in equation \Iii.
When there is no spin dependence
(Calogero-Sutherland model) $\Delta_M(u)$ can be
expressed as a determinant and
 coincides with the invariants given
in \cal .

\bigskip
Let us obtain here the eigenvalues of
 $\Delta(u)$.
Our method is first to diagonalise
the differentials $\hat D_i$. In this way we find the eigenvectors of
 $\hat \Delta(u)$.
  Then we symmetrise these eigenvectors with respect to
 $\Sigma _3$
to obtain the eigenvectors of
 $ \Delta(u)$
with the same eigenvalue.
In order to do this,
 it is convenient to
make a gauge transformation which amounts to substitute
$z_i\d_{z_i} - {\la \over 2}
\sum_{j\neq i}\frac{z_i+z_j}{z_i-z_j}$
for $z_i\d_{z_i}$ everywhere.
Hereafter, we indicate with a prime the gauged transformed quantities.
The gauge transformed differentials
 $\hat D_i'$ take the following form:
\eqn\WWi{
\hat D_i' =
 z_i\d_{z_i}
+ \la \({{M+1\over 2}-i}\)
+ \la \sum_{j>i} \theta_{ij} (K_{ij}-1)
- \la \sum_{j<i} \theta_{ji} (K_{ij}-1)  }
Their action leaves invariant the space of polynomials:
\eqn\WW{
\Psi _{[n]}(z_1,z_2,\cdots,z_M) =
z_1^{n_1}
z_2^{n_2} \cdots
z_M^{n_M}   }
where $[n]$ is a sequence of positive integers.
To a sequence $[n]$, we associate the partition $|n|$
where we arrange the $n_k$ in decreasing order.
We define an order on the partitions by saying that $|n|$
is larger than $|n'|$ if the first non vanishing difference
 $n_k - n'_k$
is positive. It follows from the same argument as in \su~
that the differentials
 $\hat D'_i$
are represented by block triangular matrices in the
basis
$\Psi _{[n]}(z_k)$.
Namely:
\eqn\WWii{
 \hat D'_i \Psi_{[n]} (z_k)=
\sum_{n'} (d_i)_{[n'][n]} \Psi_{[n']} (z_k) }
with
$ (d_i)_{[n'][n]} =0$
if $|n'|$ is larger than $|n|$.
Therefore, the eigenvalues of the differentials
 $\hat D'_i$
are given by the eigenvalues of the block matrices
on the diagonal.
Let us consider such  a block:
 $d_i^{|n|}=\({(d_i)_{[n'][n'']}}\)$
with $|n|$ a fixed partition
$|n|=(n_{1}\geq
n_{2}\geq\cdots \geq
n_{M}) $.
In the basis
 $|n_{\sigma_1},
  n_{\sigma_2},\cdots,
  n_{\sigma_M}>$,
the  $d_i^{|n|}$ take the following form:
\eqn\WZZ{
 d_i^{|n|}
 |n_{\sigma_1},
  \cdots,
  n_{\sigma_M}> = \Bigl( n_{\sigma _i} + \la ({M+1\over 2}-i
+ \sum_{j>i} X_{ij} - \sum_{j<i} X_{ji}) \Bigr)
 |n_{\sigma_1},
  \cdots,
  n_{\sigma_M}> }
where the $X_{ij}$ are defined by:
\eqn\WZZi{
X_{ij}
 |\cdots,
  n_{\sigma_i},\cdots,
  n_{\sigma_j},\cdots > =\cases{
 -|\cdots,
  n_{\sigma_i},\cdots,
  n_{\sigma_j},\cdots>
    &if\  $n_{\sigma_i}>n_{\sigma_j}$ \cr
  \ \ 0 &if\  $n_{\sigma_i}=n_{\sigma_j}$ \cr
 \ \ |\cdots, n_{\sigma_j},\cdots,
  n_{\sigma_i},\cdots>
    &if\  $n_{\sigma_i}<n_{\sigma_j}$ \cr}}
The matrices
$d^{|n|}_i$
are triangular when we order the states inside a block
by saying that $[n']$ is larger than $[n'']$ if
the last non vanishing difference $n'_k-n''_k$ is
positive. With the global order induced by this choice
the matrices
 $\hat D_i'$ are also triangular.
It follows that the eigenvalues,
$\delta _i^{[n]}$,
of $d_i^{|n|}$
are the diagonal matrix elements,
$(d_i)_{[n][n]}$.
One readily sees that the multiplets of eigenvalues,
$(\delta ^{[n]}_i)_{i=1,M}$, of
the $d_i^{|n|}$'s are all obtained by permuting the
components of the multiplet:
\eqn\toto{
(\delta _i^{|n|} ) =\({ n_i+\la (i-{M+1\over2})}\) }
As a result, the corresponding eigenvectors of $\hat D'_i$
form a degenerate set of eigenvectors of
 $\hat \Delta'(u)$ with the eigenvalue:
\eqn\tutu{
\delta ^{|n|}(u)=
\prod _{j=1}^{M}
 \({u-n_j-\la (j-{M+1\over2})}\) }
These eigenvectors form a representation of $\Sigma_1$,
isomorphic to the obvious representation of the permutations
on the sequences $[n]$.
To obtain the eigenvectors of $\Delta (u)$, one must
combine these eigenvectors with a spin component and
symmetrise the tensor product with respect to $\Sigma_3$.
The eigenvectors
associated to a given partition $|n|$
probably organise into an irreducible representation of the Yangian.

 We refer to \ref\gaud{M.Gaudin, Saclay preprint SPhT/92-158.}
for a complementary discussion of the
spectrum in the Calogero-Sutherland models.

\newsec{The spin model.}

Our understanding of the spin models\ha \sha~is less clear.
In particular, we have not yet been able to find a simple
generating function for their conserved quantities.
One can find a mapping from the $su(p)$ dynamical model at $\la =1$
 onto the $su(p+1)$
spin chain which extends the known mapping from the
Calogero-Sutherland model onto the $su(2)$ spin chain.
However, the computations are cumbersome and shed no light on
the transmutation of $su(p)$ into $su(p+1)$.
In this section, we use the transfer matrix formalism
to deduce the decomposition of the $su(2)$ spin chain into irreducible
Yangian representations. From this knowledge, one can easily
reconstruct the spectrum and its degeneracies.

The Hamiltonian of the $su(p)$ spin model is:
\eqn\Vi{ H_S =\sum_{i\not=j} h_{ij}(P_{ij}-1) }
The indices $i,j$ refer to the sites of the chain which we take
 of length $N$. The function $h_{ij}$ is the same as above,
eq. \Iii .

The transfer matrix is the limit $\la\to\infty$,
$\frac{u}{\la}=x$ fixed, of the matrix \IViii :
\eqn\Vii{
 T^0(x)= 1 +\sum_{i,j=1}^N P_{0i}\Bigl({\inv{x-L} }\Bigr)_{ij} }
with $L_{ij}=(1-\de_{ij})\te_{ij}P_{ij}$.
If we let the distance between the sites
go to infinity with $z_1<<z_2\cdots <<z_N$, the
$ \te_{ij} $ converge to the step function $ \te(i-j) $
and the transfer matrix reduces to its usual form:
\eqn\Vx{
T^0(x) =
\Bigl({ 1 +\frac{P_{01}}{x} }\Bigr)
\Bigl({ 1 +\frac{P_{02}}{x} }\Bigr) \cdots
\Bigl({ 1 +\frac{P_{0N}}{x} }\Bigr) }
For generic values of the complex numbers $z_j$,
the induced representation
of the Yangian algebra $Y(sl_p)$ is
irreducible. Thus, its quantum determinant
 is a c-number which we can
evaluate on any vector. Choosing the vector with all
spins $\sig_j$ equal to $p$ gives:
\eqn\Vvi{
 Det_q\ T(x)\ =\ 1 + \sum_{i,j=1}^N \Bigl({ \inv{x-\Theta}}\Bigr)_{ij}
 =\ \frac{ \De_N(x+1) }{\De_N(x) } }
Here $\Theta$ is the $N\times N$ matrix with matrix elements $\th_{ij}$
and $ \De_N(x) $ is its characteristic polynomial:
\eqn\Viii{
 \De_{N}(x)= det(x-\Theta) }

Although the transfer matrix satisfies
the exchange relations \IVi~ for all values of the parameters $z_i$,
it commutes with the Hamiltonian  \Vi~ only if
$\sum_j h_{ij}(\te_{ij}-\te_{ji})=0$ \nous .
 In order to satisfy this condition,
 we choose either
$z_k=\exp(i2\pi k/N)$ with $k=1,\cdots, N$,
(trigonometric models),
or, $z_k=\exp(\ga k)$, with $k$ integers,
(hyperbolic models).
Diagonalising $\Theta$, we obtain the
following expression for $\De_N$ in the trigonometric case:
\eqn\Zzi{
\De_N(x)= \prod_{j=1}^N \bigl( x-\frac{N+1}{2} +j \bigr) }

We recall that in the trigonometric case, the Yangian algebra is
completely reducible. Let us obtain here its decomposition in
 the $su(2)$ case.
It is known that irreducible representations of the $su(2)$
Yangians are characterised by a polynomial $P(x)$ \drin \ref\press
{V.Chari and A.Pressley, J.Reine Angew.Math. 417 (1991) 87. }.
Namely, in a canonical normalisation, the transfer matrix,
$T_C(x)$, is a rational function of $x$ and one can find a
state, $|\Om >$, such that:
\eqn\Zut{
T_C(x) |\Om > =
\pmatrix{
|\Om > & 0 \cr
\ast   & \frac{P(x+1)}{P(x)} |\Om >\cr }  }
In each irreducible
block appearing in the decomposition, the transfer matrix $T(x)$
differs from the canonical one only by a multiplicative
factor, $T(x)=\vphi(x)T_C(x)$.
Let us evaluate the quantum determinant of $T(x)$ in this
block:
\eqn\Zzii{
\frac{\De(x+1)}{\De(x)}=
\frac{P(x+1)}{P(x)}\vphi(x)\vphi(x-1) }
Using the fact that
 $\Delta(x) T(x)$ is polynomial in $x$, this equation implies that
the roots, $\{a_k\}_{k=1}^p$, of $P(x)$ are among those of $\De (x)$.
If we denote by
$\{ b_j \} _{j=1}^{N-p}$ the remaining roots of $\De(x)$,
it admits the solution:
\eqn\Zzv{
\varphi(x) =\prod_{j=1}^{{N-p}\over 2}\frac{(x-b_{2j-1}+1)}{(x-b_{2j-1})},
\qquad
P(x) = \prod_{k=1}^p (x-a_k) }
only if the $b_j$ come in disjoint pairs $(b_{2j-1},b_{2j}=b_{2j-1}+1)$.
Let us label the intervals separating consecutive roots of
$\De(x)$ by a 1 if the extremities of the interval are
 $ b_{2j-1},b_{2j} $ for some j, and by a 0 if not.
Since the intervals are disjoint, two 1 can never be adjacent.
If we add a $0$ at both extremeties of the sequence, we code the possible
irreducible representations appearing in the decomposition
by a sequence of $N+1$ symbols 0 and 1.
In this way, we recover the characterisation of the
spectrum in terms of $motifs$ given in \nous .
Namely, a motif is a series of $q$ consecutive $0$ bordered
by $1$'s; it corresponds to the natural spin $\frac{q-1}{2}$ representation
$ su(2) $. The representation content of a sequence is then
the irreducible tensor product of its $motifs$.
\bigskip
\noindent {\bf Acknowledgments}

We thank M.Douglas for an illuminating discussion.
\listrefs

\end